\documentclass[reqno,11pt]{amsart}

\usepackage{amsmath}
\usepackage{amssymb}
\newtheorem{Theorem}{Theorem}[section]

\newtheorem{Corollary}[Theorem]{Corollary}

\theoremstyle{definition}

\numberwithin{equation}{section}

\begin{document}

\title{On the Heat kernel and the Korteweg--de Vries hierarchy}
\dedicatory{To Pierre van Moerbeke on the occasion of his 
60th birthday}

\author[P.~Iliev]{Plamen~Iliev}
\address{School of Mathematics, Georgia Institute of Technology, 
Atlanta, GA 30332--0160, USA}
\email{iliev@math.gatech.edu}

\date{September 14, 2004}

\begin{abstract}
We give explicit formulas for Hadamard's coefficients in terms of the 
$\tau$-function of the KdV hierarchy. We show that some of the basic 
properties of these coefficients can be easily derived from these formulas. 
The first immediate corollary is the symmetry of Hadamard's coefficients 
about the diagonal. Another well known fact, which follows from this approach, 
is that on the diagonal Hadamard's coefficients determine the right-hand sides 
of the equations of the KdV hierarchy. The proof of the main result uses Sato 
theory and simple properties of Gegenbauer polynomials.
\end{abstract}

\maketitle

\newcommand{\thref}[1]{Theorem \ref{#1}}
\newcommand{\coref}[1]{Corollary \ref{#1}}
\newcommand{\cC}{{\mathcal C}}
\newcommand{\cA}{{\mathcal A}}
\newcommand{\cP}{{\mathcal P}}
\newcommand{\cR}{{\mathcal R}}
\newcommand{\cS}{{\mathcal S}}
\newcommand{\cV}{{\mathcal V}}
\newcommand{\cW}{{\mathcal W}}
\newcommand{\C}{\mathbb C}
\newcommand{\Z}{\mathbb Z}
\newcommand{\res}{\mathrm{res}}
\newcommand{\pd}{\partial}
\newcommand{\pdt}{\tilde{\partial}}
\newcommand{\expp}{\exp\left(xz+\sum_{i=1}^{\infty}s_{2i-1}z^{2i-1}\right)}
\newcommand{\expm}{\exp\left(-xz-\sum_{i=1}^{\infty}s_{2i-1}z^{2i-1} \right)}
\newcommand{\Psip}{\bar{\Psi}}
\newcommand{\Psim}{\bar{\Psi}^*}
\newcommand{\Sch}{\mathfrak{S}}

%%%%%%%%%%%%%%%%%%%%%%%%%%%%%%%%%%%%%%%%%%%%%%%%%%%%%%          Section 1

\section{Introduction and examples}

Consider the one-dimensional Schr\"odinger (or Sturm-Liouville) operator
\begin{equation}
L=\frac{\pd^2}{\pd x^2}+u(x).
\end{equation}
Its heat kernel $H(x,y,t)$ is the fundamental solution of the heat equation 
\begin{equation}\label{1.2}
\left(\frac{\pd}{\pd t}-L\right)f=0.
\end{equation}
It is well known that $H(t,x,y)$ has an asymptotic expansion of the form
\begin{equation}
H(x,y,t) \sim \frac {e^{-\frac{(x-y)^2}{4t}}}{\sqrt{4\pi t}} \left( 1 +
\sum_{n=1}^{\infty} H_n(x,y)t^n\right) \text{ as }t\rightarrow 0+.
\end{equation}
The differential equation \eqref{1.2} for $H(x,y,t)$ implies the 
recursion-differential equations for the coefficients $H_n=H_n(x,y)$:
\begin{align}
&H_0=1   \label{1.4}\\
&(x-y)\frac{\pd H_n}{\pd x}+nH_n=LH_{n-1} \text{ for }n\geq 1. \label{1.5}
\end{align}
This system is known to admit unique smooth solutions $H_n=H_n(x,y)$ in some 
neighborhood of the diagonal $x=y$. The coefficients $H_n$ are named after 
J. Hadamard \cite{Hadamard}, who constructed them for the first time.

Computation of heat invariants of self-adjoint elliptic operators is 
a well known problem in spectral theory which has many applications, in 
particular to geometry and theoretical physics 
\cite{BGV,Berger,Fulling,Gilkey,Kac,McKS}. The asymptotics of the 
one-dimensional Schr\"odinger operator are of 
particular interest due to their relations to the Korteweg-de Vries 
(KdV) hierarchy. More precisely, it is known that the restriction of the 
heat coefficients on the diagonal gives the right-hand sides of the KdV 
hierarchy, see \cite{McKvM,Sch}.

In the present paper we show that there are simple formulas for the 
Hadamard's coefficients $H_n(x,y)$ in terms of the $\tau$-function of the 
KdV hierarchy (see the next section for a precise definition of the 
$\tau$-function). 

A remarkable explicit formulas for the coefficients of the Taylor expansion 
of $H_n(x,y)$ around the diagonal $x=y$ were previously constructed in 
\cite[Theorem 1.3]{ASch}. However, these formulas have a rather complicated 
combinatorial structure and it is practically impossible to write a closed 
formula for the coefficients even for simple potentials $u$. One advantage of 
the formulas derived in this paper is that they give finite expressions for 
the heat coefficient if the $\tau$-function is known (e.g. the solitons, or 
the more general algebro-geometric solutions of KdV).

To see the importance of the KdV equations, 
let us compute the first few coefficients using the defining relations 
\eqref{1.4}-\eqref{1.5}. Anticipating the appearance of the $\tau$-function, 
let us write $u(x)$ as
\begin{equation*}
u(x)=2\frac{\pd^2 \log(\tau(x))}{\pd x^2}.
\end{equation*}
From \eqref{1.5} one can easily obtain simple formulas for $H_1$ and 
$H_2$
\begin{equation}
H_1(x,y)=\frac{2}{x-y}\left(\frac{\tau'(x)}{\tau(x)}-
\frac{\tau'(y)}{\tau(y)}\right)
\end{equation}
and
\begin{equation}
H_2(x,y)=\frac{2}{(x-y)^2}\left(\left(\frac{\tau''(x)}{\tau(x)}+
\frac{\tau''(y)}{\tau(y)}\right)-H_1(x,y)
-2\frac{\tau'(x)\tau'(y)}{\tau(x)\tau(y)}\right).
\end{equation}
For the third coefficient we have the following formula
\begin{equation}\label{1.8}
\begin{split}
&(x-y)^3H_3(x,y)=-6(x-y)H_2(x,y)
+2\left(\frac{\tau'''(x)}{\tau(x)}-\frac{\tau'''(y)}{\tau(y)}\right)\\
&\quad 
-2\left(\frac{\tau''(x)\tau'(x)}{\tau^2(x)}
   -\frac{\tau''(y)\tau'(y)}{\tau^2(y)}\right)
+4\left(\frac{\tau'(x)}{\tau(x)}\frac{\tau''(y)}{\tau(y)}
   -\frac{\tau'(y)}{\tau(y)}\frac{\tau''(x)}{\tau(x)}\right)\\
&\quad +\frac{4}{3}\left(\left(\frac{\tau'(x)}{\tau(x)}\right)^3-
\left(\frac{\tau'(y)}{\tau(y)}\right)^3\right)+\int_{y}^x u^2(\xi) d \xi.
\end{split}
\end{equation}

Notice that the integral cannot be computed explicitly, unless something 
remarkable happens. This is the place where the KdV equation comes in. 
Assume that $u(x)$ depends on a additional parameter $s_3$ and it satisfies 
the KdV equation
\begin{equation}\label{1.9}
4\pd_3 u = u'''+6uu',
\end{equation}
where $\pd_3=\pd/\pd s_3$ stands for the partial derivative with respect to 
$s_3$, and $u'$ is the derivative with respect to $x$. Then one can easily 
see that 
\begin{equation}
\begin{split}
&\int_{y}^x u^2(\xi) d \xi = -\frac{2}{3}\left(\frac{\tau'''(x)}{\tau(x)}
-\frac{\tau'''(y)}{\tau(y)}\right)
+2\left(\frac{\tau''(x)\tau'(x)}{\tau^2(x)}
-\frac{\tau''(y)\tau'(y)}{\tau^2(y)}\right)\\
&\quad -\frac{4}{3}\left(\left(\frac{\tau'(x)}{\tau(x)}\right)^3-
\left(\frac{\tau'(y)}{\tau(y)}\right)^3\right)
+\frac{8}{3}\pd_{3}\log \frac{\tau(x)}{\tau(y)},
\end{split}
\end{equation}
which combined with \eqref{1.8} leads to a simple formula for $H_3(x,y)$.

We'll extend these computations by showing that if $u$ is a solution of 
the KdV hierarchy and $\tau$ is the corresponding $\tau$-function, then there 
are simple explicit formulas for $H_n(x,y)$ in terms of $\tau$. 

The paper is organized as follows. In the next section we recall some basic 
facts about the KdV hierarchy and Sato theory, which are needed for the 
formulation and the proof of the main result. 
In Section 3, we prove a general formula for $H_n(x,y)$. It is interesting 
that the smoothness of the coefficient $H_n(x,y)$ on the 
diagonal is related to the Gegenbauer polynomials. As a corollary of the 
main theorem, we see the symmetry of the coefficients about the diagonal 
$x=y$ as well as the connection between $H_n(x,x)$ and KdV equations.

As another application of the explicit formula, we show in \cite{I} that the 
expansion is finite if and only if the potential $u(x)$ is a 
rational solution of the KdV hierarchy decaying at infinity studied in 
\cite{AM,AMM}. Equivalently, one can characterize the corresponding operators 
as the rank one bispectral family in \cite{DG}. For related results 
concerning the finiteness property of the heat kernel expansion on the 
integers and rational solutions of the Toda lattice hierarchy see \cite{GI}. 
For solitons of the Toda lattice and purely discrete versions of the heat 
kernel see \cite{Haine}.

\section{Korteweg-de Vries hierarchy and Sato theory}
In this section we recall some basic facts about KdV hierarchy and Sato 
theory. For more details on this and the more general 
Kadomtsev-Petviashvili hierarchy we refer the reader to the papers 
\cite{SS,DJKM} or the more detailed expositions \cite{Dickey,vM}.

Let 
$$L=\frac{\pd^2}{\pd x^2}+u(x)$$ 
be a second order differential operator. The KdV hierarchy is defined by the 
Lax equations
\begin{equation}\label{2.1}
\frac{\pd L}{\pd s_j}=[(L^{j/2})_+\;,L], 
\end{equation}
where $j=1,3,5,\dots$ is an odd positive integer and $(L^{j/2})_+$ is the 
differential part of the pseudo-differential operator $L^{j/2}$. 
The first equation (for $j=1$) simply means that 
$u(x,s_1,s_3,s_5,\dots)=u(x+s_1,s_3,s_5,\dots)$, giving us the convenience 
to occasionally identify $x$ and $s_1$. The next equation (for $j=3$) is 
exactly the KdV equation \eqref{1.9}. Let us represent $L$ in a dressing form 
\begin{equation}\label{2.2}
L=W\pd^2 W^{-1},
\end{equation}
where $W$ is a pseudo-differential operator of the form
\begin{equation}\label{2.3}
W=\sum_{k=0}^{\infty}\psi_k\pd^{-k}, \quad \psi_0=1.
\end{equation}
The wave (Baker) function $\Psi(x,s,z)$ and the adjoint wave function 
$\Psi^*(x,s,z)$ are defined as
\begin{equation}\label{2.4}
\begin{split}
\Psi(x,s,z)&= W \expp \\
& =\left(\sum_{k=0}^{\infty}\psi_k z^{-k}\right)\expp
\end{split}
\end{equation}
and
\begin{equation}\label{2.5}
\begin{split}
\Psi^*(x,s,z)&= (W^*)^{-1} \expm \\
& =\left(\sum_{k=0}^{\infty}\psi^*_k z^{-k}\right)\expm,
\end{split}
\end{equation}
where $W^*$ is the formal adjoint to the pseudo-differential operator $W$.
Using \eqref{2.2} one can easily see that 
\begin{equation}\label{2.6}
L\Psi(x,s,z)=z^2\Psi(x,s,z)\text{ and }L\Psi^*(x,s,z)=z^2\Psi^*(x,s,z).
\end{equation}
We shall also use the reduced wave function $\Psip$ and the reduced 
adjoint wave function $\Psim$ obtained from $\Psi$ and $\Psi^*$, respectively, 
by omitting the exponential factor, i.e. 
\begin{equation}
\Psip(x,s,z)=\sum_{k=0}^{\infty}\psi_k z^{-k}
\end{equation}
and 
\begin{equation}
\Psim(x,s,z)=\sum_{k=0}^{\infty}\psi^*_k z^{-k}.
\end{equation}
Equations \eqref{2.6} imply 
\begin{equation}\label{2.9}
L\Psip(x,s,z)+2z\pd_x \Psip(x,s,z)=0\text{ and }
L\Psim(x,s,z)-2z\pd_x\Psim(x,s,z)=0.
\end{equation}
Using equations \eqref{2.2}-\eqref{2.5} one can show that the wave and the 
adjoint wave function satisfy the following bilinear identities
\begin{equation}\label{2.10}
\res_z\left(z^{2n}\Psi^{(l)}(x,s,z)\Psi^*(x,s,z)\right)=0,
\end{equation}
for all nonnegative integers $n$ and $l$, where $\Psi^{(l)}(x,s,z)$ is the 
$l$th derivative of $\Psi$ with respect to $x$, and the residue is around 
$z=\infty$. 

The remarkable discovery of the Kyoto school was that the KdV hierarchy 
\eqref{2.1} could be described by a function $\tau(x,s)$. The 
reduced wave and the reduced adjoint wave functions can be expressed in terms 
of $\tau(x,s)$ by the following formulas
\begin{equation}\label{2.11}
\Psip(x,s,z)=\frac{\tau(x;s-[z^{-1}])}{\tau(x,s)}
\text{ and }
\Psim(x,s,z)=\frac{\tau(x;s+[z^{-1}])}{\tau(x,s)},
\end{equation}
where $[z]=(z,z^3/3,z^5/5,\dots)$.

Finally, let us denote by $W_n(x,y)$ the coefficients of the  
function\footnote{This function is closely related to the Green function for 
$L$.}\\ 
$\Psip(x,s,z)\Psim(y,s,z)$, i.e.
\begin{equation}\label{2.12}
\Psip(x,s,z)\Psim(y,s,z)=\sum_{n=0}^{\infty}W_n(x,y)z^{-n}.
\end{equation}
Using \eqref{2.11} we can easily write an explicit formula for $W_n$ in 
terms of the $\tau$-function. If we denote by $\Sch_k(s)$ the elementary 
Schur polynomials defined by 
\begin{equation}
\sum_{k=0}^{\infty}\Sch_k(s)z^k=
\exp\left(\sum_{k=1}^{\infty}s_{2k-1}z^{2k-1} \right),
\end{equation}
then we have
\begin{equation}\label{2.14}
W_n(x,y)=\frac{\sum_{k=0}^n
[\Sch_k(-\pdt)\tau(x,s)]\;[\Sch_{n-k}(\pdt)\tau(y,s)]}{\tau(x,s)\tau(y,s)},
\end{equation}
where
\begin{equation}
\pdt=(\pd_1,\pd_3/3,\dots,\pd_{2k-1}/(2k-1),\dots).
\end{equation}

\section{Explicit formulas for Hadamard's coefficients}

The main result of the paper is the following theorem.
\begin{Theorem} The Hadamard's coefficients can be computed from the following 
relation
\begin{equation}\label{3.1}
H_n(x,y)=(-1)^n\sum_{k=0}^{n-1}\frac{2^{n-k}(n-k)_{2k}}{k!}
\frac{W_{n-k}(x,y)}{(x-y)^{n+k}},
\end{equation}
where $(\alpha)_k=\alpha(\alpha+1)\dots(\alpha+k-1)$ denotes the 
Pochhammer symbol, and $W_n(x,y)$ are defined by \eqref{2.14}.
\end{Theorem}

\begin{proof}
To prove that the Hadamard's coefficients are given by \eqref{3.1} we need to 
check that \eqref{1.5} holds and that $H_n(x,y)$ are smooth on the 
diagonal $x=y$.

To see that \eqref{1.5} holds, let us denote 
\begin{equation}
f_n(x,y,z)=(-1)^n\sum_{k=0}^{n-1}
\frac{2^{n-k}(n-k)_{2k}}{k!(x-y)^{n+k}}z^{n-k-1}.
\end{equation}
Then \eqref{3.1} can be rewritten as
\begin{equation}
H_n(x,y)=\res_z\left[f_n(x,y,z)\Psip(x,s,z)\Psim(y,s,z)\right].
\end{equation}

Using the last equation together with \eqref{2.9} one can easily see that
\begin{align*}
&\left[(x-y)\pd_x+n\right]H_n(x,y)-L(x,\pd_x)H_{n-1}(x,y)\\
&\quad =\res_z\Big[
\big((x-y)\pd_x f_n(x,y,z)+nf_n(x,y,z)-\pd_x^2f_{n-1}(x,y,z)\big)
\Psip(x,s,z)\Psim(y,s,z)\\
&\qquad + 
\big((x-y)f_n(x,y,z)+2zf_{n-1}(x,y,z)-2\pd_xf_{n-1}(x,y,z)\big)
\pd_x\Psip(x,s,z)\Psim(y,s,z)
\Big].
\end{align*}

A direct computation now shows that
\begin{align*}
& (x-y)\pd_x f_n(x,y,z)+nf_n(x,y,z)-\pd_x^2f_{n-1}(x,y,z)=0\\
& (x-y)f_n(x,y,z)+2zf_{n-1}(x,y,z)-2\pd_xf_{n-1}(x,y,z)=0,
\end{align*}
which proves \eqref{1.5}.

Next we need to show that $H_n(x,y)$ is well defined on the diagonal. Writing 
$H_n(x,y)$ as
\begin{equation}
H_n(x,y)=\frac{2(-1)^n}{(x-y)^{2n-1}}
\sum_{k=0}^{n-1}\frac{2^k(x-y)^{k}(k+1)_{2n-2k-2}}{(n-k-1)!}W_{k+1}(x,y),
\end{equation}
and applying L'H\^opital's rule we see that we need to prove that for 
$j=0,1,\dots,2n-2$ we have 
\begin{equation}\label{3.5}
\sum_{k=0}^{n-1}2^k\binom{j}{k}\frac{(2n-2-k)!}{(n-k-1)!}\,\pd_x^{j-k}
W_{k+1}(x,y)|_{x=y}=0.
\end{equation}
Using \eqref{2.12} and 
$$\pd_x^j\Psip(x,s,z)=\expm (\pd_x-z)^j\Psi(x,s,z)$$
we see that \eqref{3.5} is equivalent to the following identities
\begin{equation}\label{3.6}
\res_z\left[\left(\sum_{k=0}^{n-1}2^k\binom{j}{k}\frac{(2n-2-k)!}{(n-k-1)!}\,
z^k(\pd_x-z)^{j-k}\Psi(x,s,z)\right)\Psi^*(x,s,z)\right]=0
\end{equation}

Equation \eqref{3.6} will follow from the bilinear identities \eqref{2.10} 
if we can prove that the polynomial
\begin{equation}\label{3.7}
P_{n,j}(w)=
\sum_{k=0}^{n-1}2^k\binom{j}{k}\frac{(2n-2-k)!}{(n-k-1)!}\,(w-1)^{j-k},
\end{equation}
is an even/odd function when $j$ is an even/odd number, respectively. 
It is a pleasant surprise to see that these polynomials are closely related to 
very well known classical orthogonal polynomials - the so called 
Gegenbauer polynomials.

The Gegenbauer (or ultraspherical) polynomials are defined by
\begin{equation}\label{3.8}
\cC_n^{\lambda}(w)=\sum_{k=0}^n2^k\binom{\lambda+k-1}{k}
\binom{2\lambda+n+k-1}{n-k}(w-1)^k,
\end{equation}
see for example \cite[pages 302--303]{AAR}. 
Notice that this definition can be used for arbitrary $\lambda$. If
$\lambda>-1/2$  and $\lambda\neq 0$ these polynomials orthogonal on the 
the interval $(-1,1)$ with respect to $(1-x^2)^{\lambda-\frac{1}{2}}$ which, 
in particular, implies that $\cC_n^{\lambda}(w)$ is an even/odd function 
when $n$ is even/odd, respectively. However, we need these polynomials also 
for negative values of $\lambda$. In this case, we can use the three 
term reccurence relation
\begin{equation}
2(n+\lambda)w\cC_n^{\lambda}(w)=(n+1)\cC_{n+1}^{\lambda}(w)
+(n+2\lambda-1)\cC_{n-1}^{\lambda}(w).
\end{equation}
and $\cC_0^{\lambda}=1$ and $\cC_1^{\lambda}=2\lambda w$ to 
deduce that $\cC_n^{\lambda}(w)$ is an even/odd polynomial when $n$ is 
even/odd, respectively.

Changing the summation index in \eqref{3.7} we can rewrite $P_{n,j}(w)$ as
\begin{equation}\label{3.10}
P_{n,j}(w)=
\sum_{k=\max(0,j-n+1)}^{j}2^{j-k}\binom{j}{k}\frac{(2n-j-2+k)!}{(n-j+k-1)!}\,
(w-1)^{k}.
\end{equation}
From the last equation and the defining relation \eqref{3.8} for the 
Gegenbauer polynomials one can see that\footnote{$(-1)!!=1$ and 
$(2k-1)!!=1\cdot 3\cdots(2k-1)$ for $k\geq 1$.}
\begin{align}
P_{n,j}(w) &= j!2^{n-1}(2n-2j-3)!!\;\cC_{j}^{n-j-\frac{1}{2}}(w)
&\text{ for } 0\leq j\leq n-1,\\
P_{n,j}(w) &= \frac{(-1)^{j-n+1}j!2^{n-1}}{(2j-2n+1)!!}\;
\cC_{j}^{n-j-\frac{1}{2}}(w) &\text{ for }n\leq j \leq 2n-2,
\end{align}
which completes the proof.
\end{proof}

From \eqref{2.14} and \eqref{3.1} we obtain the following

\begin{Corollary} The Hadamard's coefficients $H_n(x,y)$ are symmetric 
functions of $x$ and $y$, i.e. we have
$$H_n(x,y)=H_n(y,x).$$
\end{Corollary}

Finally, we show that the heat coefficients $\{H_n(x,x)\}$ determine the 
right-hand sides of KdV equations \eqref{2.1}.

\begin{Corollary} We have 
\begin{equation}\label{3.13}
H_n(x,x)=\frac{2^n}{(2n-1)!!}W_{2n}(x,x)
\end{equation}
and 
\begin{equation}\label{3.14}
[(L^{\frac{2n-1}{2}})_+,L]=2\pd_x W_{2n}(x,x).
\end{equation}
Thus, the KdV hierarchy \eqref{2.1} is equivalent to the following equations
\begin{equation}
\pd_{2n-1}u=\frac{(2n-1)!!}{2^{n-1}}\pd_x H_n(x,x), \text{ for } n=1,2,\dots.
\end{equation}
\end{Corollary}

\begin{proof} Using \eqref{3.1} and applying L'H\^opital's rule $2n-1$ times 
we see that
\begin{equation}\label{3.16}
\begin{split}
H_n(x,x)&=(-1)^n\frac{2}{(2n-1)!}\\
&\qquad\times\sum_{k=0}^{n-1}2^k\binom{2n-1}{k}\frac{(2n-k-2)!}{(n-k-1)!}
\pd_x^{2n-1-k}W_{k+1}(y,x)|_{y=x}\\
& = (-1)^n\frac{2}{(2n-1)!}
\res_z\left[z^{2n-1}\left(P_{n,2n-1}(z^{-1}\pd_x)\Psi(x,s,z)\right)\;
\Psi^*(x,s,z)\right],
\end{split}
\end{equation}
where $P_{n,2n-1}(w)$ is the polynomial defined by \eqref{3.10} for $j=2n-1$
\begin{equation}
P_{n,2n-1}(w)=
\sum_{k=n}^{2n-1}2^{2n-1-k}\binom{2n-1}{k}\frac{(k-1)!}{(k-n)!}\,
(w-1)^{k}.
\end{equation}
Notice that this time we have
\begin{equation}
P_{n,2n-1}(w)=
(-1)^n2^{n-1}(2n-2)!!\left(\cC^{-n+\frac{1}{2}}_{2n-1}(w)+1\right),
\end{equation}
and using the same argument (the bilinear identity and the fact that 
$\cC_{2n-1}^{-n+\frac{1}{2}}(w)$ is an odd polynomial) we obtain 
\eqref{3.13} from \eqref{3.16}.

From \eqref{2.2}, \eqref{2.4}, \eqref{2.5} and \eqref{2.11} it follows that
\begin{equation*}
L^{\frac{2n-1}{2}} = W \pd_x^{2n-1} W^{-1}= 
\sum_{i,j=0}^{\infty} \frac{\Sch_i(-\tilde{\pd})\tau(x,s)}{\tau(x,s)}\;
\pd_x^{2n-1-i-j}\cdot \frac{\Sch_j(\tilde{\pd})\tau(x,s)}{\tau(x,s)}.
\end{equation*}
Combining this formula with \eqref{2.14} we see that the coefficient of 
$\pd_x^{-1}$ in $L^{\frac{2n-1}{2}}$ is $W_{2n}(x,x)$. If we denote by 
$(L^{\frac{2n-1}{2}})_-$ the integral (Volterra) part of the 
pseudo-differential operator $L^{\frac{2n-1}{2}}$ we obtain
\begin{align*}
&[(L^{\frac{2n-1}{2}})_+,L]=[(L^{\frac{2n-1}{2}})_+,L]_+
=[L^{\frac{2n-1}{2}}-(L^{\frac{2n-1}{2}})_-,L]_+=[L,(L^{\frac{2n-1}{2}})_-]_+\\
&= [\pd_x^2+u(x),W_{2n}(x,x)\pd_x^{-1}+O(\pd_x^{-2})]_+
=2\pd_x(W_{2n}(x,x)),
\end{align*}
which gives \eqref{3.14} and completes the proof.
\end{proof}

%%%%%%%%%%%%%%%%%%%%%%%%%%%%%%%%%%%%%%%%%%%%%%%%%%%%%%          Bibliography

\end{document}